\documentclass[10pt,english,prb,amsmath,amssymb,prb,showkeys,superscriptaddress,twocolumn,showpacs,floatfix]{revtex4-2}
\usepackage{graphicx}
\usepackage{tabularx}
\usepackage{epstopdf}
\usepackage{dcolumn}
\usepackage[colorlinks=true,linkcolor=blue ,citecolor=blue,urlcolor=blue]{hyperref}
\usepackage{xr-hyper}    
\usepackage{multirow}
\usepackage[usenames,dvipsnames]{xcolor}
\usepackage{soul}
\usepackage{braket}
\usepackage{bm}
\usepackage{nicefrac}
\usepackage{siunitx}
\usepackage{color,transparent}
\usepackage[utf8]{inputenc}
\usepackage{upgreek}

\usepackage{amsmath}   
\usepackage{mathtools} 
\usepackage{amssymb}   
\usepackage{mathrsfs}  
\usepackage{textcomp}  
\usepackage{accents}   
\usepackage{bm}        
\usepackage{soul}
\usepackage{relsize}
\usepackage{graphicx}  
\usepackage{epstopdf}  
\usepackage{subfigure} 
\usepackage{makecell}
\usepackage{hhline}
\usepackage{cellspace}
\usepackage{array}     
\usepackage{tabularx}  
\usepackage{multirow}  
\usepackage{booktabs}  
\usepackage{dcolumn}   
\usepackage{gensymb}   
\usepackage{setspace}
\usepackage{scrextend}
\usepackage{float}

\begin{document}
\setlength{\heavyrulewidth}{0.08em}
\setlength{\lightrulewidth}{0.05em}
\setlength{\cmidrulewidth}{0.03em}
\setlength{\belowrulesep}{0.65ex}
\setlength{\belowbottomsep}{0.00pt}
\setlength{\aboverulesep}{0.40ex}
\setlength{\abovetopsep}{0.00pt}
\setlength{\cmidrulesep}{\doublerulesep}
\setlength{\cmidrulekern}{0.50em}
\setlength{\defaultaddspace}{0.50em}
\setlength{\tabcolsep}{4pt}

\title{DFT-Guided Operando Raman Characterization of Ni-Based Phases Relevant to Electrochemical Systems}

\newcommand{\TUDa}{Surface Science Laboratory, Department of Materials and Geosciences, Technical University of Darmstadt, Peter-Grünberg-Straße 4, 64287 Darmstadt, Germany}
\newcommand{\TUDb}{Technische Universit\"at Darmstadt, Fachbereich Material und Geowissenschaften, Fachgebiet Materialmodellierung, Otto‑Berndt‑Straße 3, 64287 Darmstadt, Germany}
\newcommand{\USC}{SmartState Center for Experimental Nanoscale Physics, Department of Physics and Astronomy, University of South Carolina, Columbia, SC 29208}
\newcommand{\TUDc}{Advanced Electron Microscopy Division, Institute of Materials Science, Department of Materials and Geosciences, Technical University of Darmstadt, Peter-Grünberg-Straße 2, Darmstadt 64287, Germany.}

\newcommand{\equalcontrib}{$^\dagger$}

\altaffiliation{These authors contributed equally to this work.}
\author{Harol Moreno Fernández}
\email{harol.moreno@tu-darmstadt.de}
\affiliation{\TUDa}

\author{Siavash Karbasizadeh}
\email{siavashk@email.sc.edu}
\affiliation{\USC}

\author{Esmaeil Adabifiroozjaei}
\affiliation{\TUDc}

\author{Leopoldo Molina-Luna}
\affiliation{\TUDc}

\author{Jan P. Hofmann}
\affiliation{\TUDa}

\author{Mohammad Amirabbasi}
\email{amirabbasi@tu-darmstadt.de}
\affiliation{\TUDb}

\begin{abstract}

We present a phase-resolved investigation of Ni-based oxides and hydroxides relevant to the oxygen evolution reaction (OER), combining ground-state DFT+$U$ calculations with operando and in situ Raman spectroscopy, supported by high-resolution TEM. Five crystalline phases—cubic and hexagonal NiO, monoclinic and trigonal Ni(OH)$_2$, and NiOOH—are systematically characterized in terms of their vibrational and electronic structure. Although the DFT models are idealized (0 $K$, defect-free, no solvation), they serve as clean, phase-specific references for interpreting complex experimental spectra. Cubic NiO is confirmed to be dynamically and electronically stable, consistent with dominant Raman modes observed experimentally. Despite dynamic instabilities in phonon dispersions, hexagonal NiO is structurally verified via TEM, suggesting substrate- or defect-stabilized metastability. Ni(OH)$_2$ polymorphs are both vibrationally stable semiconductors, with the trigonal phase exhibiting stronger spin polarization. NiOOH exhibits spin-polarized electronic states across the Brillouin zone, consistent with its asymmetric band structure under ferromagnetic ordering. Independently, phonon calculations reveal soft modes near the $\Gamma$-point, indicating dynamic instability under idealized conditions, yet operando Raman spectra align closely with calculated zone-center modes. However, introducing 0.03 \AA ~symmetry-breaking displacements relaxes the NiOOH lattice off its saddle point, removing imaginary phonon modes and stabilizing the phase. This integrated framework demonstrates how idealized DFT can reveal intrinsic fingerprints that anchor the interpretation of vibrational and electronic responses in catalytically active, dynamically evolving Ni-based materials.

\end{abstract}

\maketitle

\section{Introduction}
\label{sectionI}
Nickel-based materials remain among the most widely investigated catalysts for the oxygen evolution reaction (OER) in alkaline water splitting due to their cost-effectiveness and high redox tunability \cite{Yeo_2012,kumaravel_2022,Dionigi_2023}. These catalysts undergo dynamic transitions among NiO, Ni(OH)$_{2}$, and NiOOH, each exhibiting distinct local coordination geometries and ligand chemistries. These structural differences modulate key electronic and vibrational properties, including oxidation state stabilization, lattice dynamics, and Ni–O hybridization. For instance, in NiO, Ni$^{2+}$ ions are octahedrally coordinated by oxide anions in a dense rock salt lattice. Despite its simple structure, strong electron–electron correlations localize the Ni $3d$ electrons, resulting in a Mott insulating ground state. In contrast, Ni(OH)$_{2}$ contains hydroxyl ligands instead of oxides, enabling hydrogen bonding and imparting greater vibrational flexibility due to the presence of O–H stretching and bending modes. NiOOH, with a Ni$^{3+}$ oxidation state, has a mixed-ligand coordination environment that is structurally asymmetric and metastable but becomes electrochemically relevant under anodic polarization \cite{li_coordination_2022,Chang_2014,Romanenko_1993}. Understanding these phase transformations is essential for interpreting electrocatalytic behavior, particularly under dynamic polarization  conditions.

\begin{figure*}[]
\centering
\includegraphics[scale=0.55]{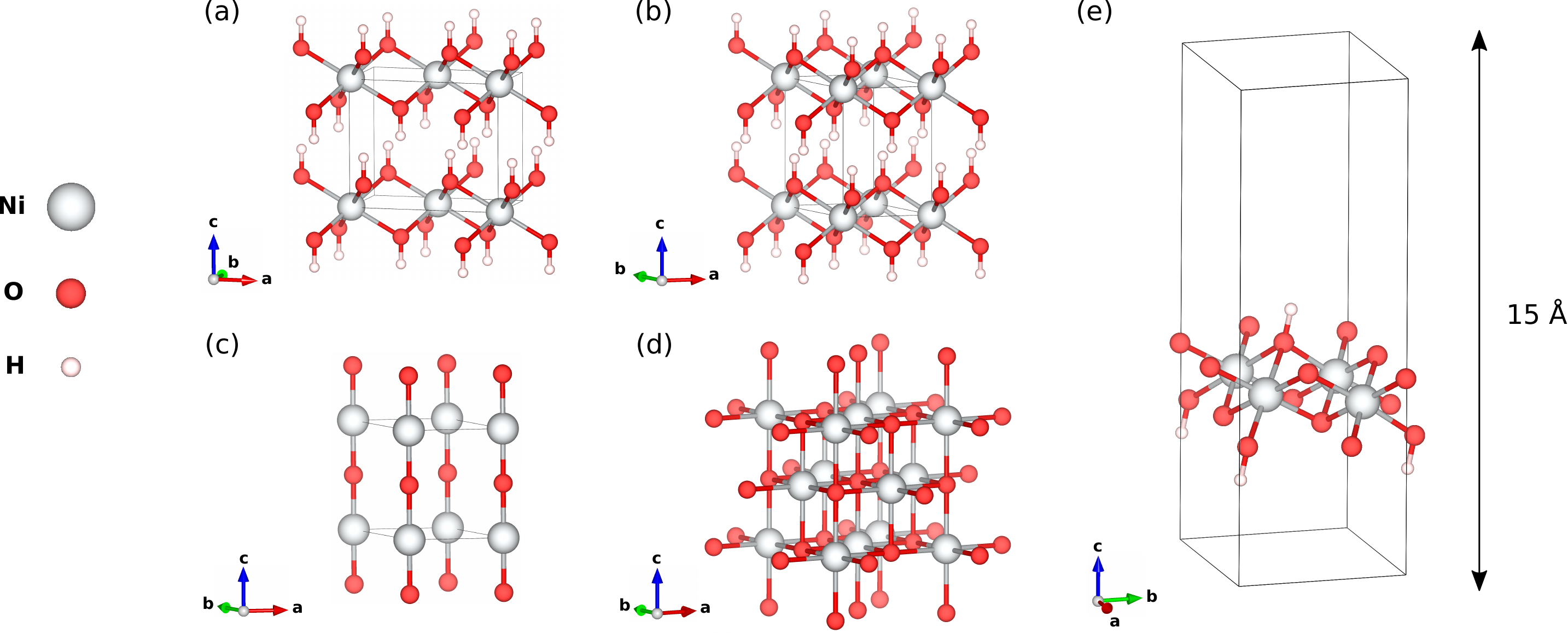}
\caption{\footnotesize Crystallographic geometry of (a) monoclinic Ni(OH)$_2$, (b) trigonal Ni(OH)$_2$, (c) hexagonal NiO, (d) cubic NiO,  and (e) NiOOH phases. The silver-gray, red, and pale-pink spheres denote Ni, O, and H atoms. The figures have been generated using VESTA~\cite{momma2011vesta}.}
\label{Fig. 1}
\end{figure*}

Raman spectroscopy, especially under operando conditions, enables real-time in situ tracking of vibrational features associated with phase transitions, hydration dynamics, and redox transitions. Hall et al. \cite{hall_raman_2012} demonstrated a clear spectroscopic distinction between $\alpha$-Ni(OH)$_2$ and $\beta$-Ni(OH)$_2$ phases using Raman and IR spectroscopy, showing how vibrational features shift with structural order, hydration, and anion incorporation. Similarly, Yeo and Bell \cite{Yeo_2012} used operando Raman to show how $\beta$-NiOOH exhibits superior OER activity compared to $\gamma$-NiOOH, linking this to distinct Raman-active modes and structural transformations induced by electrochemical cycling. Additionally, we have used in-situ Raman to track changes and verify the stability of NiOOH in KOH, allowing us to study the different kinetics associated with phase instability \cite{Moreno_2024}.

While operando Raman spectroscopy provides valuable insight into vibrational responses during electrochemical operation, it cannot fully resolve phonon dynamics or electronic structure—especially in complex, multi-phase systems. To complement the experimental data, we employed ground-state DFT+$U$ calculations of phonon dispersions and spin-resolved electronic band structures. These idealized models, though lacking solvent effects and electrochemical potentials, serve as controlled references that isolate intrinsic vibrational and electronic signatures of each phase. This approach enables direct mapping between theoretical predictions and operando Raman spectra, facilitating accurate phase identification and interpretation of dynamic behavior under applied bias. Additionally, while DFT+DMFT or hybrid functionals may better handle dynamic correlations, they are prohibitively costly for multi-phase phonon modeling. Therefore, DFT+$U$, in contrast, accurately localizes Ni 3$d$ electrons, revealing trends in vibrational stability, magnetic structure, and orbital hybridization across Ni-based phases. Despite being ground-state and defect-free, these idealized models serve a precise analytical role: they isolate phase-specific vibrational responses that can be directly mapped onto operando Raman spectra. The comparison enables confident mode assignments and detection of dynamic disorder.

Magnetic ordering fundamentally influences NiO's vibrational properties, particularly through its coupling to optical phonon modes. In its bulk form, NiO typically adopts a rocksalt-type cubic structure with well-established type-II antiferromagnetic (AFM) ordering below the Néel temperature (523 K). This magnetic structure consists of ferromagnetically aligned (111) planes that alternate antiferromagnetically, stabilized by superexchange interactions mediated through O$^{2-}$ \cite{mironova-ulmane_raman_2007}. In our study, we analyze both the conventional cubic phase and the hexagonal polymorph of NiO, as both phases may coexist under our experimental conditions. Our DFT calculations are conducted for both phases to assess how their structural and magnetic characteristics influence vibrational behavior. 

The surface coordination and spin structure of the solid-electrolyte interface likely deviate from the bulk ground state. Nevertheless, idealized ground-state DFT calculations are a rigorous reference for analyzing phonon and electronic features. Although defect-free and calculated in a vacuum, these models enable the clear identification of Raman-active modes and allow for a systematic comparison of vibrational responses across phases. This integration with operando Raman facilitates the interpretation of vibrational changes and structural disorders that emerge during electrochemical operation.

Aytan et al. \cite{Aytan_2017} demonstrated using UV Raman spectroscopy and DFT calculations that antiferromagnetic (AFM) ordering in NiO leads to pronounced shifts in longitudinal optical (LO) and transverse optical (TO) phonon energies—specifically, a hardening of the LO mode and a softening of the TO mode. These changes reflect strong spin–phonon coupling, with extracted coupling constants of +14.1 cm$^{-1}$ and –7.9 cm$^{-1}$, respectively, effects that are absent in non-magnetic calculations. Building on this, our study explicitly incorporates AFM ordering within the DFT+$U$ framework for both phonon dispersion and electronic band structure calculations. This approach enables a more accurate account of magnetic contributions to NiO’s vibrational and electronic properties. Notably, the computed vibrational features show strong agreement with operando Raman spectra, reinforcing the validity of our model and its utility in tracking phase evolution under applied electrochemical potentials.

The integration with operando Raman spectroscopy is not intended to replicate the exact electrochemical environment but rather to anchor the interpretation of experimentally observed vibrational changes. The theoretical phonon modes act as baseline markers against which experimental shifts, broadening, or mode activations can be mapped. By correlating theoretical phonon modes with operando Raman spectra, we can interpret spectral changes associated with structural distortions, oxidation-state transitions, and phase change under electrochemical conditions. While more realistic models include finite-temperature effects, solvation, or applied potential which could offer deeper insight, such approaches remain computationally prohibitive for systematic phonon analysis across complex, multi-phase system. Our use of idealized ground-state DFT thus provides a practical yet informative reference framework.

The organization of this paper is structured as follows: The manuscript opens with section \ref{sectionI}, which outlines the motivation for linking ground-state density-functional theory with operando Raman spectroscopy across five Ni-based oxides phases relevant to the oxygen-evolution reaction. Section \ref{section_II} traces the spectroscopic and computational fingerprints of (A) hexagonal and cubic NiO, (B) monoclinic and trigonal Ni(OH)\textsubscript{2}, and (C) NiOOH. Section \ref{sectionIII} distills the main insights on phase stability, soft-mode behavior, and catalytic implications. A standalone Methods section compiles all experimental and computational protocols. The paper closes with Appendix~\ref{app-A}, deriving the Hubbard~$U$, and Appendix~\ref{app-B}, detailing the soft-phonon stabilization of NiOOH.
\section{Results and Discussion}
\label{section_II}
\begin{figure} [!htbp]
\centering
\includegraphics*[scale=0.5]{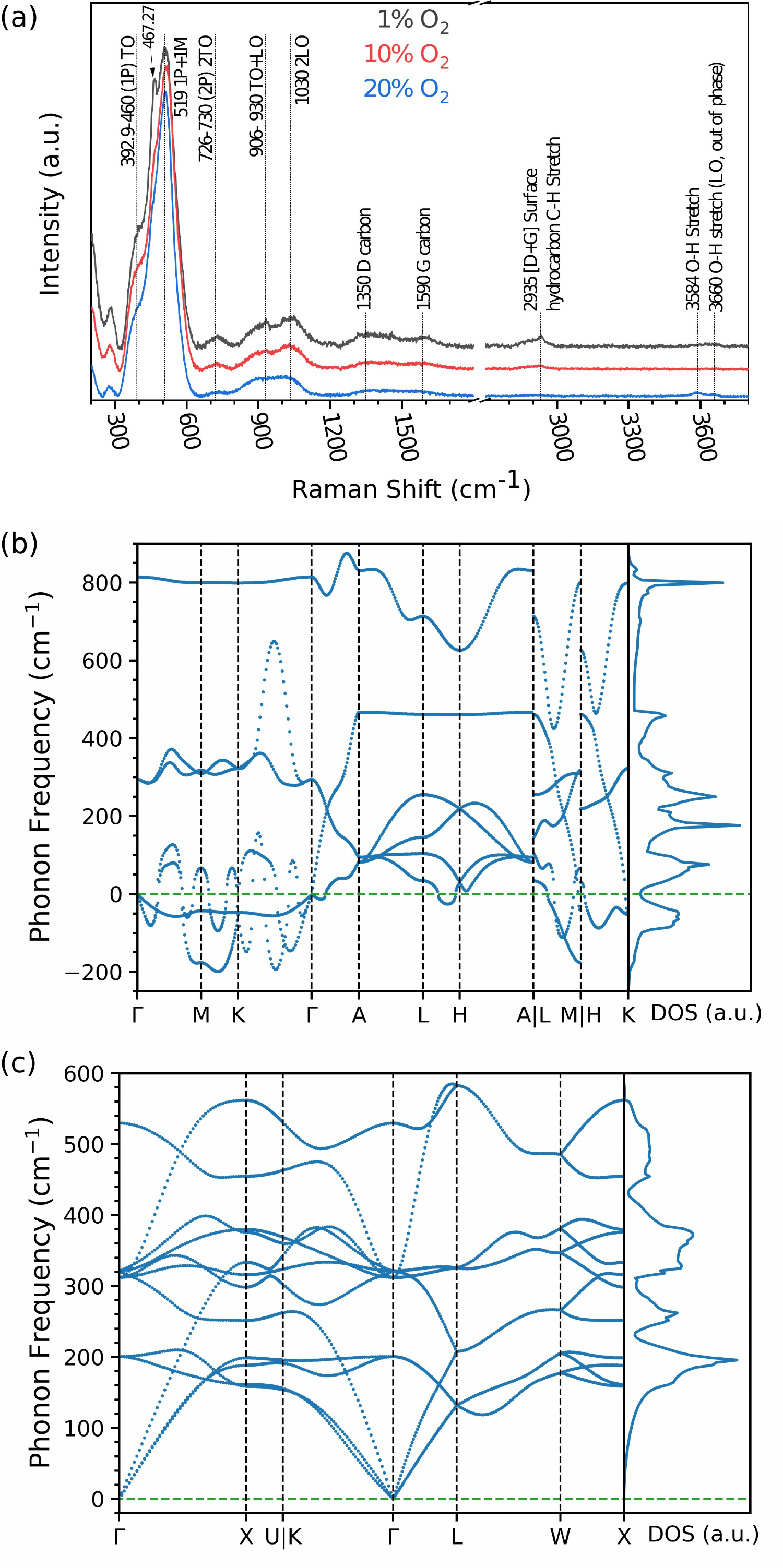}
\caption{\footnotesize\textbf{Vibrational fingerprints and lattice 
dynamics of Hexagonal \(\mathrm{NiO}\)}  
\textbf{(a)} \textit{In-situ} Raman spectra of NiO samples prepared with 1\,\%, 10\,\% and 20\% \(\mathrm{O_2}\) in the plasma discharge. It shows characteristic vibrational modes of NiO, surface carbon (D and G bands), and O–H stretching vibrations. Increasing the oxygen partial pressure broadens and attenuates selected peaks, indicating disorder induced by oxygen‑rich defects. DFT+$U$ phonon dispersion and phonon density of states (DOS) (phonon-DOS) for \textbf{(b)} hexagonal NiO and \textbf{(c)} cubic NiO. Imaginary branches (frequencies \(<0\)) along the \(\Gamma\!-\!M\), \(\Gamma\!-\!A\) and \(A\!-\!L\) directions signal dynamical instability of the ideal hexagonal lattice, consistent with the need for strain or substrate stabilization.  The calculated phonon-DOS shows intense modes at 300–500 cm\(^{-1}\), mirroring the Raman‑active region in panel (a).}
  \label{FIG-2}
 \end{figure}

\begin{figure*}[!htbp]
    \centering
    \includegraphics[scale=0.45]{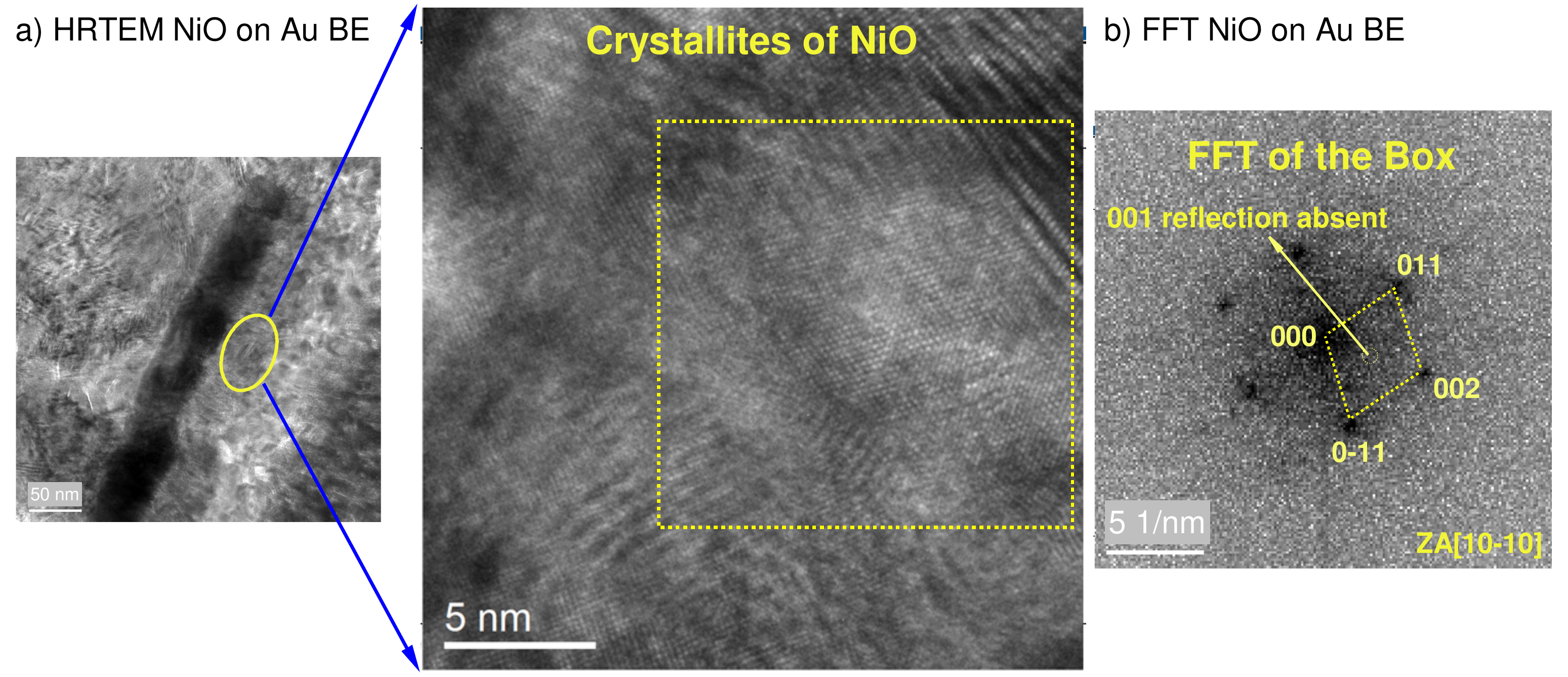}
    \caption{\footnotesize (a) HRTEM analysis at the NiO deposited on Au (b), FFT (Fast Fourier Transform) analysis providing a detailed crystallographic assessment of NiO deposited on gold (Au).}
    \label{TEM}
\end{figure*}

\begin{figure*} [!htbp]
\includegraphics[scale=0.5]{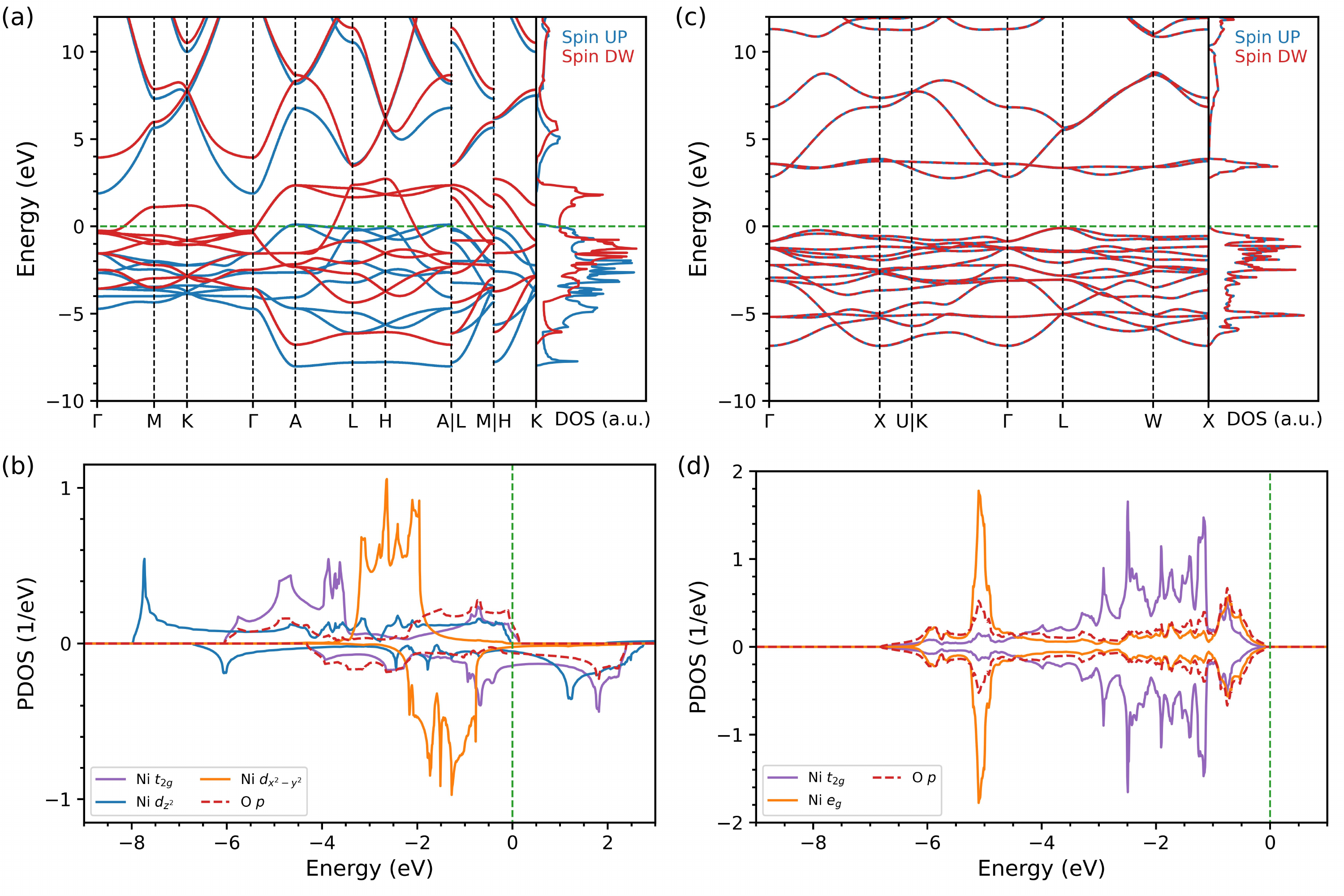}
\caption{\footnotesize
Spin‑polarized electronic band structures (left) and atom‑/orbital‑projected density of states (right) for layered nickel hydroxides of (a) hexagonal NiO, and (b) cubic NiO. Partial density of states (PDOS) is also shown for (c) hexagonal NiO, and (d) cubic NiO.
The Fermi level is set to \(E_{\mathrm F}=0\)eV (green dashed line).}
  \label{FIG-3}
 \end{figure*}

Fig.~\ref{Fig. 1} presents the crystallographic geometries of various nickel-based phases relevant to the study. In the visual representations, Silver-gray spheres represent Ni atoms, red spheres denote O atoms, and pale-pink spheres indicate H  atoms.
\subsection{Hexagonal and Cubic NiO}
\subsubsection*{\textbf{Raman, phonon band structure and DOS analysis of Hexagonal and Cubic NiO}}
Fig.~\ref{FIG-2}a shows the  Raman spectra along with the phonon band structures and phonon-DOS for hexagonal (Fig.~\ref{FIG-2}b) and cubic (Fig.~\ref{FIG-2}c) NiO. For the hexagonal phase, the phonon dispersion of the fully optimized structure still reveals several branches with imaginary frequencies, particularly along the $\Gamma$-M, $\Gamma$-A, and M-H directions, indicating that this polymorph is dynamically unstable under ambient conditions. These directions in reciprocal space correspond to important crystallographic orientations. $\Gamma$-M and M-H paths lie within the basal plane of the hexagonal structure, probing lattice dynamics within the Ni–O layers. Imaginary modes along these directions suggest in-plane structural instabilities, such as soft phonon modes, bond-angle distortions, or shear deformations. In contrast, $\Gamma$-A direction extends along the c-axis, perpendicular to the Ni–O layers, and instabilities here point to weak interlayer interactions, stacking faults, or out-of-plane acoustic mode softening\cite{zellagui_2024}.

In contrast, the phonon dispersion of cubic NiO (Fig.~\ref{FIG-2}c) reveals no imaginary frequencies across the Brillouin zone, confirming its dynamic stability at 0 K. The corresponding phonon-DOS displays sharp, well-defined peaks in the 300–600 cm$^{-1}$ region, in close agreement with experimental Raman spectra. These include prominent transverse optical (TO) modes between 370–450 cm$^{-1}$, consistent with one-phonon TO vibrations (1P–TO) \cite{uchiyama_2010,Aytan_2017}, and a distinct longitudinal optical (LO) mode near 520 cm$^{-1}$, which aligns with the sharp experimental Raman peak at 519 cm$^{-1}$, further supported by its temperature-independent spectral behavior and comparative analyses \cite{Mironova_2019}. This strong correspondence between the calculated vibrational features and experimental observations provides compelling evidence that the dominant Raman-active modes originate from the cubic NiO phase, reinforcing its structural and dynamic stability under the measured conditions. Spectra collected from NiO thin films on both Ni (from this study) and Au (from our previous study\cite{gallenberger_stability_2023}) substrates exhibit identical features, suggesting that any potential substrate-induced structural differences are not spectroscopically resolved under our measurement conditions. However, high-resolution TEM of NiO grown on Au (Fig.~\ref{TEM}) reveals the presence of a hexagonal NiO polymorph localized at the metal support-catalyst interface—likely limited to the first few nanometers. This interfacial phase is not expected to contribute measurably to the Raman signal. Its structural identification nonetheless supports the inclusion of hexagonal NiO in our DFT analysis, providing insight into the vibrational and electronic behavior of this metastable phase and its potential role in local interface phenomena.

Higher-order Raman features such as bands around 726–730 cm$^{-1}$ (assigned to 2TO overtones), 906–930 cm$^{-1}$(TO + LO combinations) \cite{Dietz_1971,Dietz_1971_2}, and a broader band at $\sim$1030 cm$^{-1}$ (2LO) are attributed to multi-phonon scattering processes, as confirmed by both experimental and theoretical studies \cite{Lacerda_2017,Mironova_2019,Pressl_1996}. In the high-frequency region, weak features at $\sim$2930 cm$^{-1}$ correspond to C–H stretching vibrations from surface hydrocarbons \cite{Snyder_1978,Margaret_1977}, while the sharp peaks at 3584 cm$^{-1}$ and 3660 cm$^{-1}$ are assigned to O–H stretching modes, with the latter representing a longitudinal optical (LO) out-of-phase vibration of surface hydroxyl groups \cite{Edwards_2006,George_book}. Though not described in harmonic phonon calculations, these modes reflect surface chemistry and hydration effects.

The dominant first- and second-order Raman-active features align well with the vibrational modes predicted for cubic NiO, supporting its dynamic and structural stability under the conditions studied. In contrast, the hexagonal NiO phase, while sharing some overlapping Raman signatures, displays dynamic instabilities at \(0~\text{K}\) in DFT-based phonon dispersion—evidenced by imaginary modes along both in-plane ($\Gamma$-M, M-H)) and out-of-plane ($\Gamma$-A) directions. These results suggest that hexagonal NiO is intrinsically metastable and would require external stabilization, such as strain, defects, or substrate interactions.

The combined application of in situ Raman spectroscopy, ground-state DFT phonon calculations, and high-resolution TEM confirms that while hexagonal NiO can form under specific conditions, the measured vibrational RAMAN response in our study is dominated by the cubic phase. HRTEM and FFT analyses (Fig.~\ref{TEM}) reveal long-range hexagonal ordering and inter-planar spacings consistent with a hexagonal lattice, with missing reflections (e.g., 001, 010) pointing to defect-induced disorder. However, these structural defects disrupt lattice order and could play a role in stabilizing metastable polymorphs, as studied in another system by Solomon et al \cite{solomon_atomic-scale_2025}.

Notably, the Raman spectra recorded at 1$\%$, 10$\%$, and 20$\%$  O$_2$ reveal variations in peak intensity and sharpness, highlighting the role of oxygen stoichiometry in modulating the lattice vibrational spectrum. For example, higher O$_2$ content tends to reduce the intensity of peaks like 462.7 and 726 cm$^{-1}$, likely due to increased structural disorder or suppression of specific phonon modes caused by oxygen-rich defects or surface oxidation.

\subsubsection*{\textbf{Electronic band structure and DOS of Hexagonal and Cubic NiO}}

The spin-polarized band structure of hexagonal NiO (Fig.~\ref{FIG-3}a) exhibits a clear exchange splitting between the majority and minority spin channels, indicating long-range ferromagnet (FM) spin ordering. This is contrasted by cubic NiO (Fig.~\ref{FIG-3}b), where strong electron correlations and Ni 3$d$ orbital interactions stabilize antiferromagnetic (AFM) ordering, as widely demonstrated in DFT+$U$ and DFT+DMFT studies \cite{Sawatzky_1984,Leonov_2024}.
While spin polarization is included in our model of hexagonal NiO (Fig.~\ref{FIG-3}a), explicit AFM ordering—known to influence phonon energies in cubic NiO (Fig.~\ref{FIG-3}b) is not imposed. This choice may limit our ability to capture magnetic-symmetry-driven phonon effects observed in Raman spectra, particularly those arising from AFM spin–phonon coupling as shown by Aytan et al. \cite{Aytan_2017}. Although spin–phonon coupling can also occur in FM systems, we did not directly evaluate such effects in this study.

PDOS plots for both phases further illustrate these distinctions (Fig.~\ref{FIG-3}c for hexagonal, and Fig.~\ref{FIG-3}d for cubic NiO). In both cases, the valence band (VB) is primarily composed of O $2p$ and Ni $3d$ hybridized states just below the Fermi level. In contrast, the conduction band (CB) arises predominantly from unoccupied Ni $3d$ states, consistent with photoemission and DFT calculations \cite{Tjenberg_1995, Mckay_1984,KIM_1974}. For hexagonal NiO, the total DOS confirms a metallic character, with no visible band gap and significant Ni $3d$ states crossing the Fermi level. This reflects delocalized electronic behavior, where reduced Ni–O coordination in the layered structure promotes in-plane overlap of $e_{g}$ orbitals, in contrast to the Mott insulating nature of the cubic (rocksalt) phase.

In hexagonal NiO, magnetic coupling is primarily driven by direct exchange between neighboring Ni$^{2+}$ ions, whose partially filled $e_{g}$ orbitals—$d_{x^{2}-y^{2}}$ and $d_{z^{2}}$—point toward each other. Each Ni$^{2+}$ adopts the high‑spin $3d^{8}$ configuration ($t_{2g}^{6}e_{g}^{2}$), resulting in two unpaired $e_{g}$ electrons that contribute to the local magnetic moment. The significant orbital overlap between $e_{g}$ states favors ferromagnetic alignment via Hund’s-rule exchange, which favors parallel spin alignment and hence ferromagnetic coupling.

By contrast, in the cubic (rock-salt) phase, the dominant Ni-Ni interaction mechanism is superexchange mediated by O~2$p$ orbitals. This virtual hopping mechanism promotes antiferromagnetic ordering.
\subsection{Monoclinic and Trigonal Ni(OH)$_{2}$}

\subsubsection*{\textbf{Phonon band structure of monoclinic Ni(OH)$_{2}$ and trigonal Ni(OH)$_{2}$}}

Experimentally, subjecting the catalyst to 1200 cycles in 1 M KOH between 1.0 and 1.5 V (vs RHE) transforms the initial NiO substrate phase into Ni(OH)$_{2}$. This structural conversion is a necessary precursor for the subsequent formation of NiOOH, as we have previously reported \cite{Moreno_2024_1, Moreno_2024}. However, despite this phase transition, we did not observe characteristic Ni(OH)$_{2}$ vibrational modes in the Raman spectra—at least not at intensities comparable to those reported in the literature \cite{hall_raman_2012, kostecki_electrochemical_1997}. We attribute this absence to the inherently low Raman scattering cross section of Ni(OH)$_{2}$, which likely renders its vibrational modes undetectable under our experimental conditions.

\begin{figure} [!htbp]
  \centering
\includegraphics*[scale=0.57]{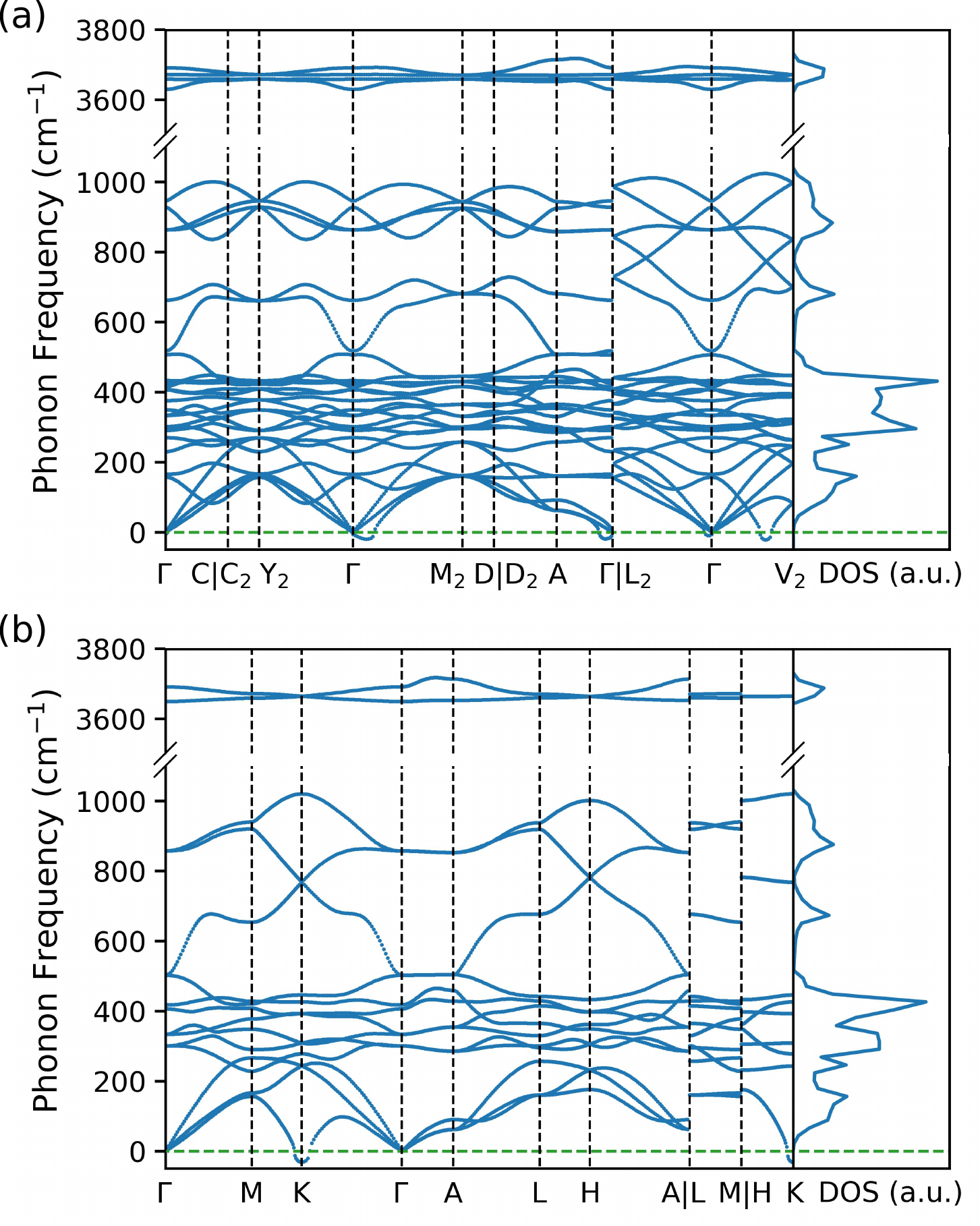}
    \caption{\footnotesize Calculated phonon dispersion relations (left) and corresponding phonon-DOS (right) for (a) monoclinic and (b) trigonal \(\mathrm{Ni(OH)_2}\).
The monoclinic and trigonal phases show no imaginary frequencies throughout the Brillouin zone, confirming dynamical stability. Nearly dispersion‑less branches at 3600–3700\,cm\(^{-1}\) arise from localized O–H stretching vibrations, while the dense manifold below 1000\,cm\(^{-1}\) is dominated by Ni–O and Ni–OH lattice modes.}
  \label{FIG-4}
 \end{figure}

However, we have included the calculated phonon dispersion and phonon-DOS for both monoclinic and trigonal Ni(OH)$_{2}$, shown in Figures ~\ref{FIG-4}a and ~\ref{FIG-4}b, respectively. As seen in Fig.~\ref{FIG-4}a, the phonon spectrum of monoclinic Ni(OH)$_{2}$ contains no imaginary frequencies across the Brillouin zone, confirming the dynamic stability of this phase at \(0~\text{K}\). This indicates that small perturbations are unlikely to trigger spontaneous structural distortions, consistent with experimental reports of its stability under ambient conditions \cite{Maryam_2023, Moreno_2024}.

The phonon dispersion reveals two well-defined vibrational regions. In the high-frequency range around $\sim$3600-3700 cm$^{-1}$, nearly flat and dispersionless branches are observed, characteristic of localized O–H stretching vibrations within the hydroxyl groups. The flatness of these modes indicates weak coupling between the light hydrogen atoms and the heavier Ni–O lattice framework \cite{Duffy_1995}. In the lower-frequency region, below 1000 cm$^{-1}$, a dense collection of modes appears, primarily associated with Ni–O stretching ($\sim$450–550 cm$^{-1}$) and Ni–OH bending ($\sim$300–400 cm$^{-1}$) vibrations. These spectral features are consistent with Raman and infrared observations reported for layered nickel hydroxides \cite{hall_raman_2012,hall_2014}. The phonon-DOS shows sharp peaks in both vibrational regions, supporting the presence of a well-ordered lattice with low intrinsic disorder. Together, these vibrational characteristics corroborate the structural integrity of monoclinic Ni(OH)$_{2}$.

\begin{figure*} [!htbp]
\includegraphics[scale=0.55]{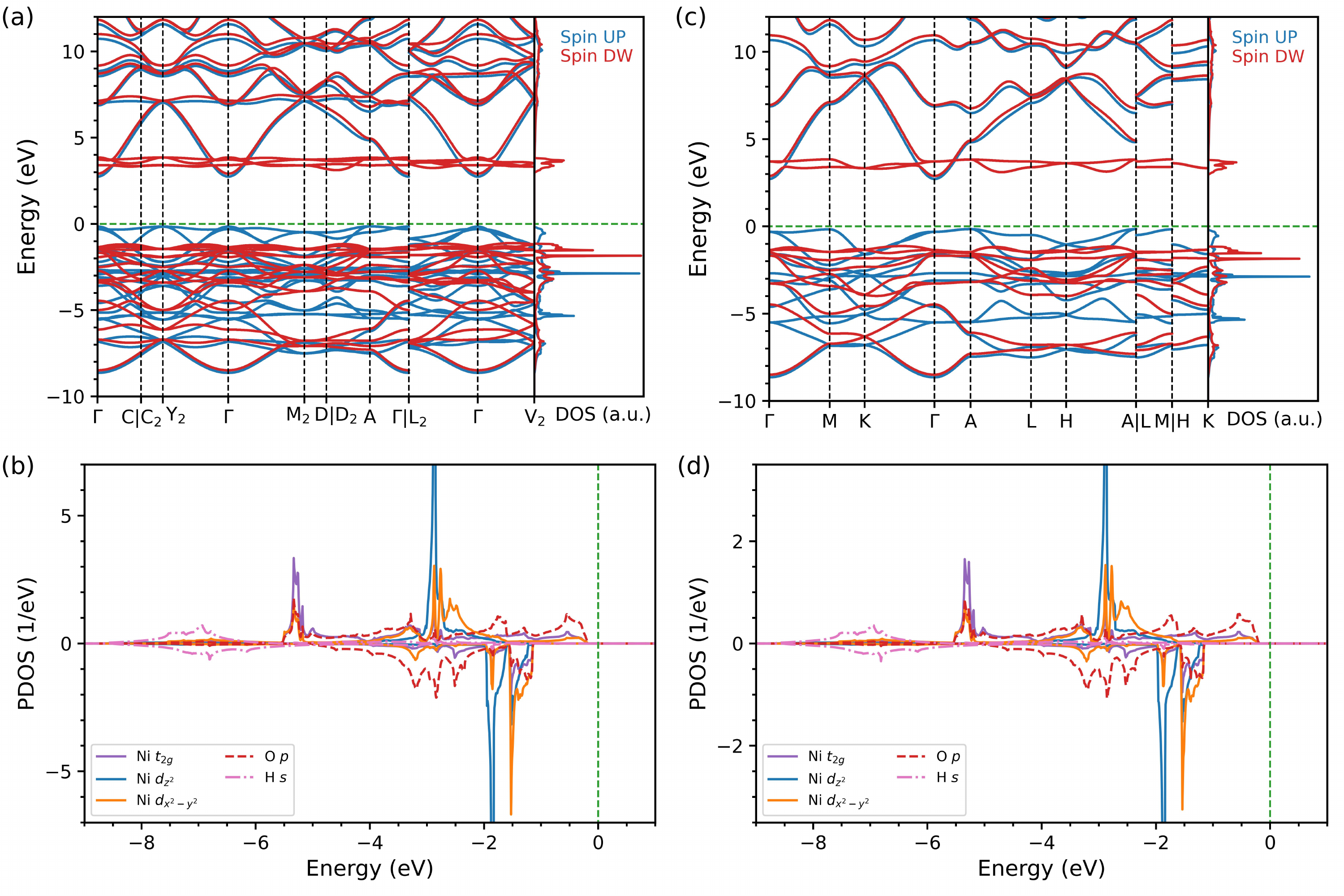}
   \caption{\footnotesize Spin‑polarized electronic band structures (left) and atom‑/orbital‑projected density of states (right) for layered nickel hydroxides of (a) Monoclinic \(\mathrm{Ni(OH)_2}\) and Trigonal \(\mathrm{Ni(OH)_2}\). Monoclinic \(\mathrm{Ni(OH)_2}\) features an indirect band gap of \(\sim 2.2~\text{eV}\) (PBE\,+\,$U$) with modest exchange splitting, reflecting the weak spin polarization of high‑spin \(\mathrm{Ni}^{2+}\,(3d^{8})\) ions. Trigonal \(\mathrm{Ni(OH)_2}\) exhibits a narrower gap of \(\sim 1.8~\text{eV}\) and a more pronounced exchange splitting, indicating stronger spin polarization and an increased propensity for magnetic ordering. PDOS is also shown for (c) Monoclinic \(\mathrm{Ni(OH)_2}\) and (d) Trigonal \(\mathrm{Ni(OH)_2}\).
In both polymorphs the VB is dominated by hybridized Ni\(3d\)–O\(2p\) states, while the CB is largely Ni\(3d\) in character.
The Fermi level is set to \(E_{\mathrm F}=0\)eV (green dashed line).
}
  \label{FIG-5}
 \end{figure*}

\begin{figure*} [!htbp]
\includegraphics[scale=0.57]{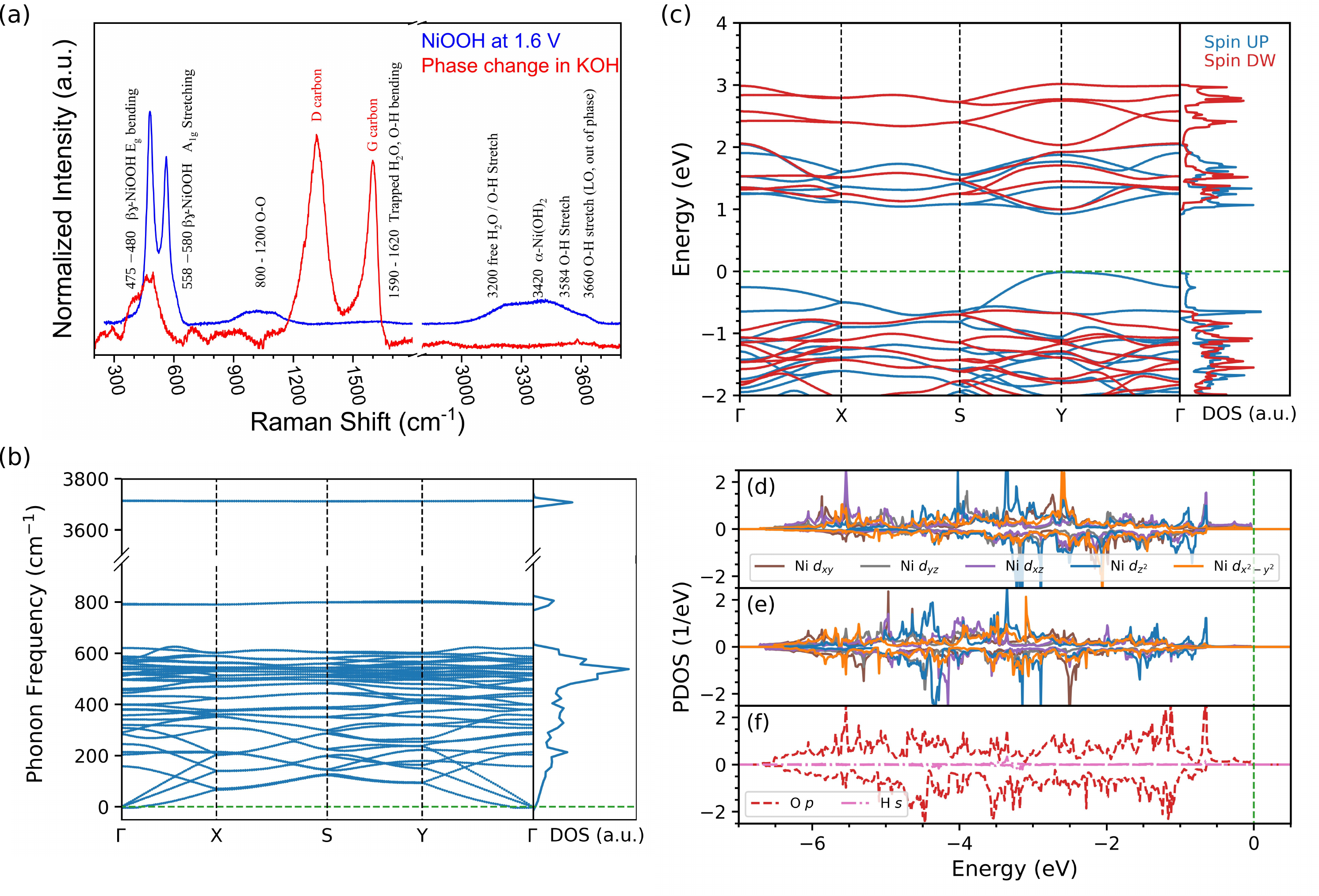}
   \caption{\footnotesize vibrational fingerprints and lattice dynamics of NiOOH (a) Operando and insitu Raman spectra of electrochemically generated NiOOH at 1.5 V vs\ RHE in 1 M KOH (blue line). It highlights the characteristic Ni–O bending and stretching modes, vibrational features from O–O species, and signals associated with trapped water and O–H stretching from free and surface-bound hydroxyl groups. In red, the phase decays after 30 minutes in  KOH at open circuit potential. After decaying, the final phase resembles the NiO from Fig. 2a.
(b) DFT‑calculated phonon dispersion and phonon-DOS for stoichiometric \(\gamma\)-NiOOH. The zone‑centre optical modes computed at \(480\) and \(560~\mathrm{cm^{-1}}\) coincide with the experimental \(E_g\) and \(A_{1g}\) peaks, and the high‑frequency manifold at 3500–3700 \(\mathrm{cm^{-1}}\) reproduces the O–H stretching region. Together, experiment and theory provide a consistent picture of NiOOH vibrational dynamics under electrochemical applied potential. (c) Spin‑polarized electronic band structure (left) and atom‑/orbital‑projected density of states (right) is shown, while parts (d, e and f) give PDOS for better identification of states for Ni high-spin, Ni low-spin, O and H, respectively.}
  \label{FIG-6}
 \end{figure*}

 Fig.~\ref{FIG-4}b shows that the trigonal polymorph also lacks imaginary phonon frequencies, indicating that this ideal structure is also stable at \(0~\text{K}\). The phonon spectrum is qualitatively similar to that of the monoclinic phase, with two main vibrational regions. The O–H stretching modes remain centered around ($\sim$ 3600–3700 cm$^{-1}$), and appear similarly sharp in both phases, reflecting localized vibrations with minimal dispersion. Below 1000 cm$^{-1}$, the phonon dispersion of the trigonal phase shows a similarly dense set of lattice vibrational modes as the monoclinic phase, including Ni–O stretching and Ni–OH bending modes. The dispersion characteristics are comparable between the two polymorphs, indicating similar interatomic force constants within the metal hydroxide layers. The phonon-DOS of the trigonal phase also features distinct peaks in this region, suggesting a well-ordered vibrational structure. These results confirm that, despite structural differences, both phases exhibit comparable lattice dynamics and dynamic stability at \(0~\text{K}\).

\subsubsection*{\textbf{ Electronic band structure of Monoclinic and trigonal Ni(OH)$_{2}$}}
Figures ~\ref{FIG-5}a, and ~\ref{FIG-5}c present the electronic band structures of monoclinic and trigonal Ni(OH)$_{2}$, respectively. Both phases exhibit clear semiconducting behavior, with indirect band gaps of  2.87$\pm$0.01 eV. In the monoclinic phase (Fig.~\ref{FIG-5}a), the VB maximum and CB minimum occur at different k-points, confirming the indirect nature of the gap. The valence‑band maximum (just below \(E_{\mathrm{F}}\)) is relatively flat near the top, indicative of localized states. At the same time, the conduction‑band minimum (just above \(E_{\mathrm{F}}\)) shows greater dispersion near the CB minimum, implying relatively lower electron effective mass and potentially higher conduction-band mobility. The trigonal phase (Fig.~\ref{FIG-5}c) also features an indirect band gap and similarly flat valence and CBs, suggesting comparable degrees of electron localization. Overall, both phases show insulating character, and no bands cross the Fermi level in either case.

Figures ~\ref{FIG-5}b, and ~\ref{FIG-5}d show the spin-resolved, PDOS for monoclinic and trigonal Ni(OH)$_{2}$, respectively. In both structures, the VB is composed primarily of hybridized O \(2p\) and Ni \(3d\) states, while unoccupied Ni \(3d\) orbitals dominate the CB. The l-resolved PDOS (~\ref{FIG-5}b) for the monoclinic phase confirms this assignment, with O \(2p\) states contributing most strongly near the VB maximum and Ni \(3d\) states spanning both the upper valence and lower CBs. The presence of H \(1s\) states deep in the valence region reflects strong localization of protons in hydroxyl groups. The low total DOS at the Fermi level and the clear separation between occupied and unoccupied states confirm that pristine monoclinic Ni(OH)$_{2}$ is a wide-gap semiconductor and remains electronically insulating unless extrinsic carriers are introduced.\\\\
A small but consistent exchange splitting between spin-up and spin-down channels is observed in both phases, originating from localized \(S=1\) moments on high‑spin \(\text{Ni}^{2+}\) (\(3d^{8}\)) ions. These local moments are captured using DFT+$U$ with an on-site Coulomb interaction $U$ of 4 eV. Although Ni(OH)$_{2}$ is paramagnetic at room temperature \cite{zhao_synthesis_2014, miyamoto_magnetic_1966}, the ferromagnetic spin alignment adopted here serves as a computationally practical approach to probe local magnetic behavior within the zero-temperature DFT framework. Interestingly, the trigonal phase shows a slightly elevated DOS near the Fermi level compared to the monoclinic form, potentially reflecting enhanced orbital overlap or subtle structural flexibility that could influence reactivity or magnetic interactions.

\subsection{NiOOH}
\subsubsection*{\textbf{Operando Raman, phonon band structure and DOS analysis of NiOOH}}
After forming a Ni(OH)$_{2}$ phase, a constant potential of 1.5 V (vs RHE) is applied for 10 minutes to form the NiOOH phase, while keeping the potential, the Raman signal is collected. Fig.~\ref{FIG-6}a (blue line) presents the operando Raman spectrum revealing vibrational features characteristic of NiOOH. Two dominant peaks at $\sim$480 cm$^{-1}$ and $\sim$560 cm$^{-1}$, corresponding to lattice vibrational modes of Ni–O \cite{diaz-morales_importance_2016}, attributed to E$_{g}$ bending and A$_{1g}$ stretching modes of $\beta$-NiOOH and $\gamma$-NiOOH\cite{Yeo_2012, bantignies_2008,desilvestro_1986}, confirming the formation of the NiOOH active phase in oxygen evolution reaction (OER).

Frequency bands in the 800-1200 cm$^{-1}$ range are attributed to O–O vibrational modes of active oxygen species, such as superoxide or peroxide intermediates \cite{diaz-morales_importance_2016}. These vibrational modes have been experimentally identified by Lee et al. \cite{Seunghwa_2023}. Additionally, broad bands in the 1590-1620 cm$^{-1}$ region are indicative of H–O–H bending vibrations \cite{hall_nickel_2015}, typically associated with trapped water molecules or adsorbed hydroxyl groups on the catalyst surface. High-frequency bands between 3200-3700 cm$^{-1}$ correspond to O–H stretching modes of hydroxyl groups within the NiOOH lattice, rather than free water, supporting the presence of structural –OH intrinsic to the oxyhydroxide. The same Fig.~\ref{FIG-6}a (red line) shows the instability of this phase, as it decays while it is immersed in KOH at open circuit potential (OCP). The spectrum is collected in situ after 30 minutes at OCP. Notice that no Ni(OH)$_{2}$ vibrations were observed experimentally; therefore, the phase change spectrum in KOH (red line) corresponds to a NiO phase, coming from the NiO bulk catalyst that was unchanged during the electrochemical process, similar signal to the one shown Fig.~\ref{FIG-2}a.

Fig.~\ref{FIG-6}b shows the calculated phonon band structure and phonon-DOS for NiOOH, providing theoretical insight into its complex vibrational behavior, which is governed by its mixed oxide-hydroxide bonding environment. This environment contributes to both lattice oxygen and O-H vibrational modes inherent to its oxyhydroxide structure. In the low‐frequency region ($\sim$ 200–650 cm$^{-1}$), the optical branches correspond to Ni–O bending and stretching modes, which directly correlate with the strong Raman bands observed at $\sim$ 350–600 cm$^{-1}$ in our experimental spectra (Fig.~\ref{FIG-6}a). Even more, around 800 cm$^{-1}$, a distinct band appears matching the Raman feature between $\sim$ 800–1200 cm$^{-1}$. At the high frequency end, a pronounced peak around $\sim$ 3700 cm$^{-1}$ in the phonon-DOS represents O–H stretching modes \cite{Yun_2021,bantignies_2008,kostecki_electrochemical_1997} matching again the intrinsic hydroxyl groups in the NiOOH lattice, visible in the operando Raman spectrum (Fig.~\ref{FIG-6}a). Despite the absence of explicit water molecules in the simulation, these modes capture the internal vibrations of bound–OH groups and explain the observed experimental features between $\sim$3200–3700 cm$^{-1}$.

Fig.~\ref{FigS2}a shows the phonon dispersion of the as-relaxed NiOOH orthorhombic structure, revealing small imaginary modes (below 0 cm$^{-1}$) near the $\Gamma$ point, which could be indicative of local dynamical instabilities in the structure\cite{pallikara_physical_2022}. This behavior aligns with our in situ Raman spectroscopy study \cite{Moreno_2024}, which demonstrated the dynamic rearrangements and phase polymorphism occurring as NiOOH decays into the more thermodynamically stable $\beta$-Ni(OH)${_2}$ phase, upon removal of the polarization potential necessary for maintaining the NiOOH active phase. These imaginary modes, therefore, further confirm that NiOOH is a metastable phase, stable only under vacuum or electrochemical operating conditions, as previously shown experimentally \cite{gallenberger_stability_2023}. 

It is important to note that the experimentally observed phase exists at finite temperatures, whereas our DFT calculations are inherently performed at $0$ K. Under this approximation, the presence of soft phonon modes reflects dynamic instability. To emulate high phase temperatures within our $0$ K DFT+U calculations, we imposed orthorhombic symmetry and manually displaced every ion by 0.03 \AA from its equilibrium position, then re‐optimized only the ionic coordinates. As shown in Fig.~\ref{FigS2}b, this procedure yields a relaxed geometry with a lower total energy than the undistorted cell. Phonon calculations on this displaced‐then‐relaxed structure, as presented in ~\ref{FIG-6}b, reveal that all imaginary modes have vanished, confirming the dynamic stability of the modified orthorhombic phase.

\subsubsection*{\textbf{Electronic band structure of NiOOH}}
Fig.~\ref{FIG-6}c presents the spin-polarized electronic band structure and DOS for stoichiometric $\gamma$-NiOOH, modeled in the ground state with all Ni atoms in the +3 oxidation state. Both spin channels exhibit a finite band gap with no states crossing the Fermi level, confirming that NiOOH behaves as a semiconductor rather than a metal or half-metal. The calculated band gap lies between 1.0–1.5 eV, slightly lower than reported (1.5–1.8 eV) \cite{Carpenter_1989, Li_structure_2021, Niklasson_2007}. 

In our analysis of NiOOH, two distinct local spin moments are observed for Ni$^{+3}$ ions: approximately 0.98 $\mathrm{\mu_{B}}$ and 0.02 $\mathrm{\mu_{B}}$. The former is consistent with a high-spin Ni$^{3+}$ configuration, indicative of one unpaired electron and a localized magnetic moment close to 1 $\mathrm{\mu_{B}}$. In contrast, the latter corresponds to a nearly quenched magnetic moment, suggesting a low-spin Ni$^{3+}$ state with fully paired electrons. This disparity in spin moments implies the presence of inequivalent Ni sites in the structure, likely arising from strong electronic correlations, local structural distortions, or charge disproportionation mechanisms. The coexistence of high- and low-spin Ni$^{3+}$ ions points to a complex electronic ground state in NiOOH, which may be driven by subtle variations in crystal field environments or enhanced covalency effects with the surrounding oxygen ligands \cite{varignon_complete_2017}.

The VB is primarily composed of O $2p$ and Ni $3d$ orbitals. In contrast, the CB consists mostly of unoccupied Ni $3d$ states, with minor contributions from H $1s$ orbitals, reflecting the oxyhydroxide nature of the structure. Fig.~\ref{FIG-6}d showing the PDOS, reveals an exchange splitting near the Fermi level, indicating the presence of localized magnetic moments on Ni, consistent with the expected high-spin Ni$^{3+}$ configuration and potential for ferromagnetic coupling.

The electronic band structure of $\gamma$-NiOOH exhibits pronounced anisotropy in the band dispersion across high-symmetry directions of the Brillouin zone. Specifically, the valence and CBs display greater curvature along the in-plane paths ($\Gamma$–M–K) and out-of-plane (A–L–H) directions, reflecting direction-dependent orbital overlap within the layered NiO$_6$ octahedral lattice, which could influence charge mobility under applied bias. Although the ground-state band structure exhibits a clean gap, the broad CBs near the CB minimum suggest that electron delocalization could be activated under electrochemical potential. Such behavior is consistent with e$_g^*$ band broadening observed in operando studies during OER \cite{Zhong_key_2023}, which has been linked to enhanced redox activity and the onset of polaronic conduction mechanisms in transition metal oxides \cite{Karsthof_2019}.

While the ground-state model does not fully reflect the dynamic conditions present during OER, it nonetheless reveals intrinsic anisotropies in the band structure. It suggests a latent capacity for direction-dependent charge transport and electronic flexibility under applied bias.

This discrepancy underscores the importance of interpreting surface-sensitive measurements in conjunction with bulk-sensitive techniques, particularly when examining metastable or crystallographically diverse oxide phases. Such materials often exhibit pronounced differences between surface and bulk regions due to variations in coordination environments, defect distributions, and local thermodynamic stability, leading to distinct structural and electronic properties across spatial domains\cite{kaviani_surface_2022, alabadleh_oxide_2003}.
\section{Conclusion}
\label{sectionIII}
In this work, we present a phase-resolved analysis of NiO-derived materials relevant to the oxygen evolution reaction, combining ground-state DFT+$U$ calculations with operando and in situ Raman spectroscopy. Despite the idealized nature of the DFT models—defect-free, fully ordered, and at $0$ K—we demonstrated that the computed vibrational and electronic properties serve as reference points, enabling meaningful correlation with the Raman response of structurally complex, electrochemically conditioned catalysts. By systematically dissecting the signatures of NiO, Ni(OH)${_2}$, and NiOOH across distinct crystallographic configurations, we have developed a consistent framework that correlates different NiO phases relevant to the OER and clarifies their metastability.

This framework revealed that cubic NiO is both dynamically and electronically stable under ambient conditions, consistent with its dominant presence in experimental Raman spectra. Hexagonal NiO, although intrinsically metastable due to phonon-mode instabilities at $0$ K, can be experimentally prepared through substrate interactions or defect-induced stabilization. Both monoclinic and trigonal Ni(OH)$_{2}$ polymorphs show robust dynamical stability and semiconducting behavior, with the trigonal phase exhibiting enhanced spin polarization, suggesting a greater potential for magnetic or redox tuning. In the case of NiOOH, our phonon dispersion analysis of the fully relaxed structure reveals soft modes near the $\Gamma$ point, indicating latent dynamic instabilities. These results, along with the previously observed phase decay under open-circuit conditions, confirm its metastable nature. However, by emulating finite-temperature conditions through atomic displacements followed by relaxation, we obtained a dynamically stable orthorhombic configuration, whose vibrational features closely match operando Raman spectra of NiOOH. This demonstrates that NiOOH can attain structural stability under applied electrochemical potentials, reaffirming its relevance as an active phase during the oxygen evolution reaction (OER).

\vspace{1.5em}  

\section*{Methods}
\subsection*{Computational Details}

\begin{table*}[!htbp]
\caption{\footnotesize
Lattice constants and space groups of the structures studied. 	
}
\begin{tabular*}{\textwidth}{c @{\extracolsep{\fill}} cc}
\hline
\hline
Material & Lattice Constants (\AA) & Space Group\\
\hline
Cubic NiO               & $a$ = 5.145 & Fm-3m \\
Hexagonal NiO           & $a$ = 2.948, $c$ = 3.432 & P6/mmm \\
Monoclinic Ni(OH)$_{2}$ & $a$ = 5.476, $b$ = 3.162, $c$ = 4.583 & C2/m \\
Trigonal Ni(OH)$_{2}$   & $a$ = 3.161, $c$ = 4.593 & P-3m1 \\
NiOOH                   & $a$ = 5.770, $b$ = 4.951 & P1 \\
\hline
\hline
\end{tabular*}
\label{T1}
\end{table*}
First-principles calculations are performed within the framework of density functional theory (DFT~\cite{DFT-1, DFT-2}) using the projector augmented wave (PAW~\cite{PAW-1, PAW-2}) method as implemented in the Vienna ab initio Simulation Package (VASP~\cite{VASP}). The exchange correlation energy is described by the generalized gradient approximation (GGA) of the Perdew-Burke-Ernzerhof (PBE) functional~\cite{PBE}. The valence electron configurations considered in the PAW pseudopotentials are $3p^{6}3d^{8}4s^{2}$ for Ni and $2s^{2}2p^{4}$ for O. 

To account for electron correlation effects within the Ni $3d$ orbitals, we employ the spherically averaged DFT+$U$ method with an on-site Coulomb interaction U of 4 eV. The on-site Coulomb parameter \( U \) for Ni $3d$ states is also calculated using the linear response method as implemented in \textsc{VASP}. The obtained \( U \) values for various Ni-containing phases are 5.86 eV, 8.65 eV, 5.63 eV, 5.18 eV, and 6.37–6.49 eV (corresponding to two inequivalent Ni sites), in the cubic, hexagonal, monoclinic, trigonal, and NiOOH phases, respectively. Although the linear response approach provides a systematic framework for determining \( U \), it does not always capture experimental material properties with sufficient accuracy. 
To ensure better agreement with experimental reults, we therefore adopt a uniform \( U \) value of 4.0 eV for all subsequent calculations, as it yields phonon band structures more consistent with experimental observations. 

A plane-wave cutoff energy of 520 eV is used, and the convergence criterion for ionic relaxation is set to a maximum force of 0.01 eV/\text{\AA}. For geometry optimization of the primitive unit cell, the Brillouin zone is sampled using the following $\Gamma$-centered $k$-point meshes \cite{KMESH} (with the corresponding structure in parentheses): 8$\times$8$\times$8 (cubic NiO), 10$\times$10$\times$10 (hexagonal NiO), 6$\times$10$\times$8 (monoclinic Ni(OH)$_{2}$), 10$\times$10$\times$8 (trigonal Ni(OH)$_{2}$), and 8$\times$8$\times$1 (NiOOH). The lattice constants and symmetries of materials considered are shown in Table~\ref{T1}. We then proceed to construct respective supercells of 3$\times$3$\times$3, 6$\times$6$\times$6, 3$\times$5$\times$4, 5$\times$5$\times$4, and 5$\times$5$\times$1 for phonon calculations, where dynamical matrices are obtained using the finite displacement method as implemented in the PHONOPY package~\cite{phonopy1, phonopy2}. The phonon calculations are conducted using $\Gamma$-point sampling for the supercells.
\subsection*{Sample preparation}   
Ni and Au discs of 12 mm diameter were polished using 15 $\mu$m and 8 $\mu$m silicon carbide (Starcke, Struers) and 0.5 $\mu$m alumina paste (MicroPolish Alumina, Buehler). NiO thin films of $\sim$100 nm were prepared by DC magnetron sputtering using a Ni target (Kurt J. Lesker with purity 99.99\%). Two mass flow controllers (MKS) were used to introduce 1.0, 2.0, and 4.0 sccm of O$_{2}$ and 19.0, 18.0, and 16.0 sccm of Ar (Air Liquide with purity 99.995\% and 99.999\%, respectively). The working pressure was set to 3 $\times$ 10$^{-2}$ mbar, and the plasma power to 15 W, with a substrate-to-target distance of 20 cm.
\subsection*{Raman Spectroscopy}
Raman spectra were acquired with a LabRAM Horiba HR-800 Raman microscope with a 633 nm laser. The power density at the sample was 1.9 mW at a spot size of 1.5 $\mu$m using a 50× long working distance objective. A 600 grooves per mm grating was chosen, the diameter of the pinhole was set to 200 $\mu$m, and the entrance slit size to 200 $\mu$m. Per spectrum, 13 scans of 10 s acquisition time were accumulated every 2.1 min. An Asymmetric Least Squares (ALS) background subtraction was applied in the specified regions of interest to facilitate the plotting and interpretation of the spectra.
\subsection*{Electrochemical Measurements}
Electrochemical measurements were conducted in a three-electrode setup in a PECC-2 cell from Zahner Elektrik, controlled by a potentiostat from Gamry Instruments (Interface 1000E). All measurements were performed in 1 M KOH (Carl Roth, unpurified, Fe $\leq$ 7ppb by ICP-OES). Before the measurement, a Hg/HgO reference electrode from ProSense was calibrated against a reversible hydrogen electrode (RHE) (HydroFlex, Gaskatel). The counter electrode was a platinum wire. By an electrochemical impedance measurement at OCP, the current response during cyclic voltammetry (CV) was manually iR corrected in the data post-treatment. The conditioning consisted of 1200 voltammetry cycles in the 1.0 V - 1.5 V (vs RHE) range with a scan rate of 100 mV s$^{-1}$, to form Ni(OH)$_2$. Additionally, 3 CVs before and after conditioning were measured between 1.0 V and 1.9 V (vs RHE) at a scan rate of 50 mV s$^{-1}$ to verify the redox wave behavior. 
Electrochemical measurements for in-situ Raman were performed using a TSC Raman cell from rhd instruments. It was operated in a three-electrode configuration. A gold-plated ring was utilized as the counter electrode at the base, while a leak-free micro Ag/AgCl electrode from Innovative Instruments served as the reference electrode calibrated against a reversible hydrogen electrode (RHE, HydroFlex, Gaskatel). Potential control was executed using a Gamry Instruments potentiostat (Interface 1000E). The NiOOH phase was formed in the film by polarizing it using chronoamperometric measurements at 1.5 V (vs RHE) for 10 min (corresponding to a potential above the Ni$^{2+}$/Ni$^{3+}$  oxidation wave).
\subsection*{Transmission electron microscopy (TEM)}
A TEM lamella of the interface of NiO deposited on Au was prepared for TEM investigation using a plasma-focused ion beam system (Helios 5 Hydra DualBeam PFIB-SEM, Thermo Fisher). The TEM analysis was done using a JEOL JEM F2100 TEM equipped with an EDS detector (X-Max 80 SDD-Detektor, Oxford)
\section*{Data Availability}
Data files for DFT calculations are available on Novel Materials Discovery (NOMAD) database using \href{https://doi.org/10.17172/NOMAD/2025.07.06-1}{DOI: 10.17172/NOMAD/2025.07.06-1}.

\section*{Author Contributions}
\textbf{Harol Moreno Fernández}: Writing – original draft, Methodology, Investigation, conceptualization, analysis, review, and editing. \textbf{Siavash Karbasizadeh}: Simulation, review, and editing. \textbf{Mohammad Amirabbasi}: Simulations, writing, coordination, review, and editing. \textbf{Esmaeil Adabifiroozjaei}: TEM characterization for NiO on Au. \textbf{Leopoldo Molina Luna}: Funding acquisition. \textbf{Jan P. Hofmann}: Funding acquisition.

\section*{Acknowledgements}
HM and JPH acknowledge the German Federal Ministry of Education and Research (BMBF) for providing financial support within the H2Giga cluster project PrometH2eus (Fkz: 03HY105H). EA and LML acknowledge the financial support of DFG in the framework of the Collaborative Research Centre Transregio 270 (CRC-TRR 270) project No. 405553726, Subproject Z01.
MA acknowledges financial support from the Collaborative Research Center FLAIR (Fermi level engineering applied to oxide electroceramics), funded by the German Research Foundation (DFG) under Project-ID No. 463184206–SFB 1548 (project A02).

\section{}
\begin{figure*}[!htbp]
\centering
\includegraphics[scale=0.5]{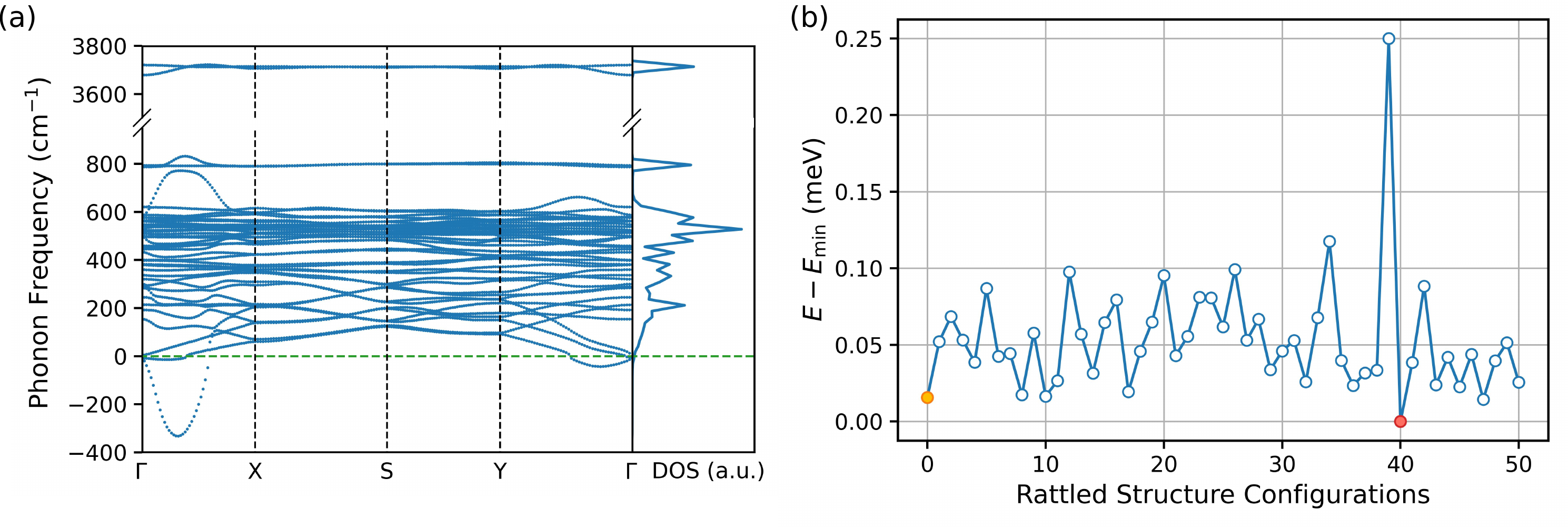}
\caption{\footnotesize (a) Phonon band structure and phonon-DOS of the initial structure of NiOOH, containing soft modes. (b) The difference between energies of different rattled structures and the structure with the lowest energy as a function of different rattled structure configurations. The golden circle shows the initial relaxed structure, while the red circle shows the structure with the lowest energy.}
\label{FigS2}
\end{figure*}

\begin{figure*}[!htbp]
\centering
\includegraphics[scale=0.5]{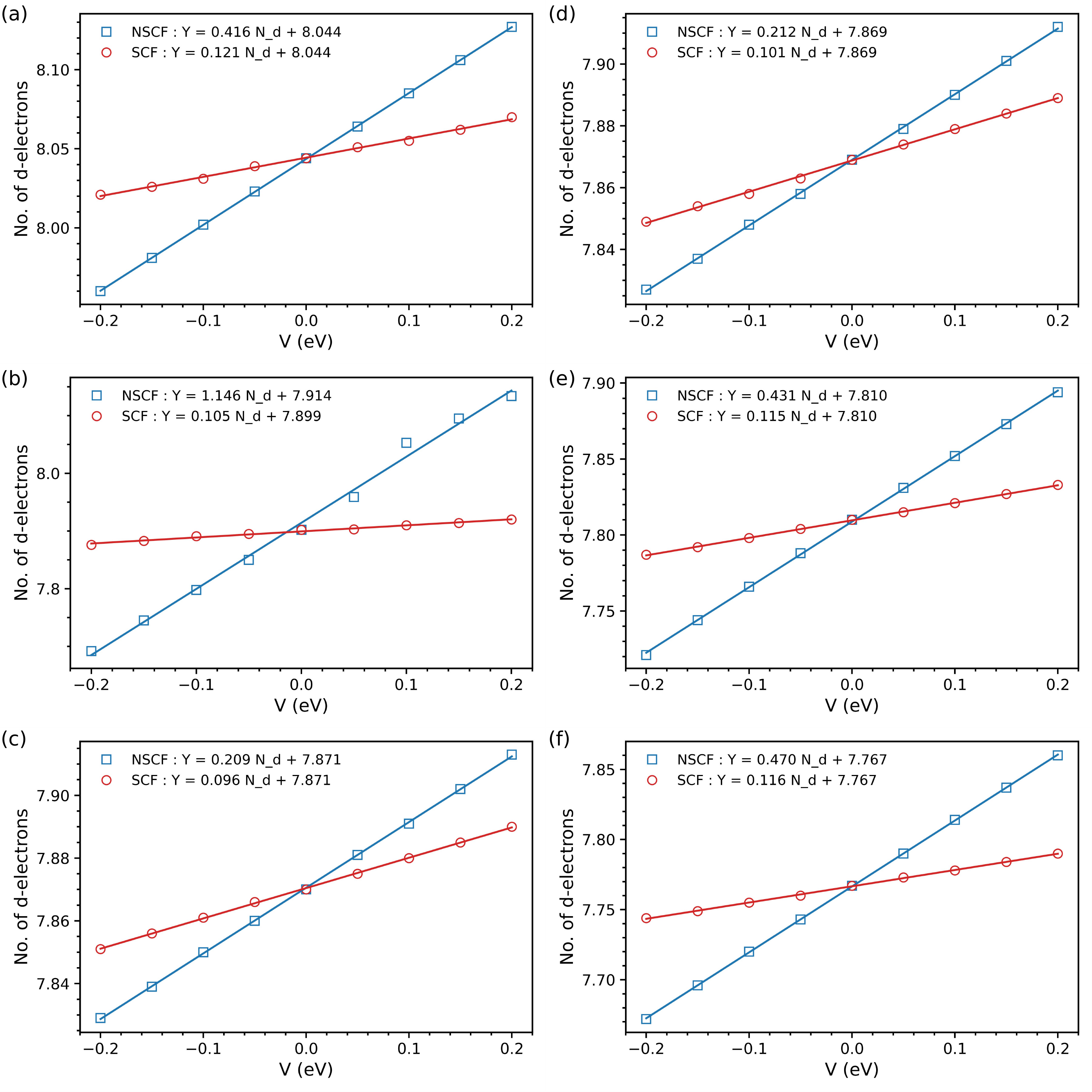}
\caption{\footnotesize Number of $d$-electrons for the singled out Ni atom as a function of added spherical potential for (a) cubic NiO, (b) hexagonal NiO, (c) Monoclinic Ni(HO)$_2$, (d) Ni(HO)$_2$, and (e), (f) NiOOH (two inequivalent Ni sites). The linear fit applied for both SCF and NSCF calculations is given in the legend.}
\label{FigS1}
\end{figure*}


\appendix
\section{\textit{Ab initio} determination of the Hubbard parameter}
\label{app-A}

The on-site Coulomb repulsion for localized electrons, commonly referred to as the 
$U$ parameter, can be calculated \textit{ab initio} approach using the linear-response method~\cite{PhysRevB.71.035105}, as implemented in the VASP code. A $2\times2\times1$ supercell is used for NiOOH, and $2\times2\times2$ supercells are employed for all other structures. $K$-point meshes of $2\times2\times2$, $1\times2\times1$, and $2\times2\times1$ are used for the hexagonal, monoclinic, and trigonal phases, respectively, while the $\Gamma$-point is used for both the cubic and NiOOH phases. Self-consistent field (SCF) and non-self-consistent field (NSCF) calculations are performed by applying an additional spherical potential to the $d$-manifold of a selected Ni atom. The resulting change in the number of $d$-electrons on this atomic site, relative to the DFT ground state, is recorded for both SCF and NSCF calculations and plotted as a function of the applied spherical potential in Fig. \ref{FigS1}. The $U$ value is then obtained from the following equation:

\begin{equation}
U=\left(\frac{\partial N_I^\mathrm{SCF}}{\partial V_I}\right)^{-1}-\left(\frac{\partial N_I^\mathrm{NSCF}}{\partial V_I}\right)^{-1},
\end{equation}
where $(\frac{\partial N_I}{\partial V_I})$ is the slope of the linear fit applied to the change in number of $d$-electrons vs. the additional spherical potential. The line equations are shown as the legend for each plot. This results in $U$ values of 5.86 eV, 8.65 eV, 5.63 eV, 5.18 eV, and 6.37–6.49 eV (corresponding to two inequivalent Ni sites), in the cubic, hexagonal, monoclinic, trigonal, and NiOOH phases.
\section{Effect of ionic displacement on soft modes in the NiOOH phase}
\label{app-B}

We investigate how displacing atomic positions from their ideal Wyckoff sites affects the soft phonon modes in the NiOOH phase. Fig. \ref{FigS2}a shows the phonon dispersion for the unperturbed high-symmetry structure (orange circle in Fig. \ref{FigS2}b), which exhibits imaginary frequencies characteristic of dynamic instabilities. To explore the local potential-energy landscape, we apply random displacements of 0.03~\AA to all atoms and subsequently relax the structure. The resulting configuration (red circle in Fig. \ref{FigS2}b) has a lower total energy and, importantly, no imaginary phonon modes, as discussed in the main text. This indicates that the original structure corresponds to a saddle point on the potential-energy surface, and that small symmetry-breaking perturbations can stabilize the lattice by lifting the soft-mode instabilities~\cite{pallikara_physical_2022}.


\bibliography{bib.bib}

\begin{figure*}[]
  \centering
  \includegraphics[width=1\textwidth]{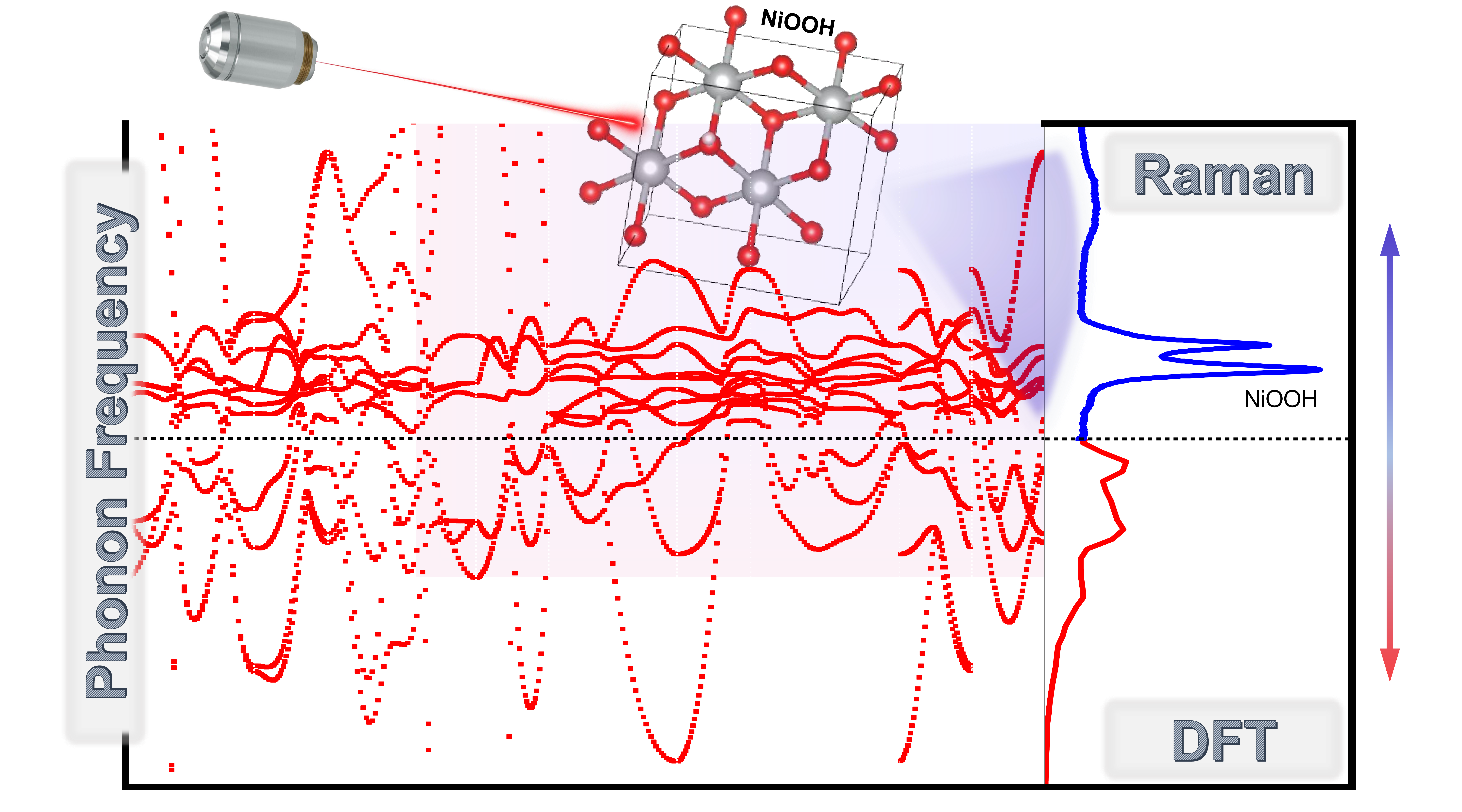}
  \caption{\footnotesize Graphical abstract}
  \label{}
\end{figure*}
\end{document}